\documentclass[preprint2]{aastex}
\usepackage{lineno}
\usepackage{amssymb}
\usepackage{ctable}
\usepackage{morefloats}
\usepackage[colorlinks,breaklinks,citecolor=blue]{hyperref}
\usepackage{graphicx,epsfig,fancyhdr,rotating,amsmath,epsf,txfonts,natbib,epstopdf,multirow,appendix}

\def \kms {{ \rm km\;s$^{-1}$}}
\def \arcsec {$^{''}$}
\def \siiv {Si\,{\sc iv}}
\def \cii {C\,{\sc ii}}

\def \si {S\,{\sc i}}
\graphicspath{{figs/}}
\begin{document}
\title{Cool transition region loops observed by the \textit{Interface Region Imaging Spectrograph}}
\author{Zhenghua Huang, Lidong Xia, Bo Li}
\affil{Shandong Provincial Key Laboratory of Optical Astronomy and Solar-Terrestrial Environment,\\ Institute of Space Sciences, Shandong University, Weihai, 264209 Shandong, China\\ \textit{z.huang@sdu.edu.cn}}
\author{ Maria S. Madjarska}
\affil{Armagh Observatory, College Hill, Armagh BT61 9DG, N. Ireland}

\date{Received date, accepted date}

\begin{abstract}
We report on the first \textit{Interface Region Imaging Spectrograph (IRIS)} study of cool transition region loops. This class of loops has received little attention in the literature, mainly due to instrumental limitations. A cluster of such loops was observed on the solar disk in active region NOAA11934, in the \siiv\,1402.8\,\AA\ spectral raster and 1400\,\AA\ slit-jaw (SJ) images. We divide the loops into three groups and study their dynamics and interaction. The first group comprises relatively stable loops, with 382--626\,km cross-sections. Observed Doppler velocities are suggestive of siphon flows, gradually changing from $-10$\,\kms\ at one end to 20\,\kms\ at the other end of the loops. Nonthermal velocities from 15\,\kms\ to 25\,\kms\ were determined. These physical properties suggest that these loops are impulsively heated by magnetic reconnection occurring at the blue-shifted footpoints where magnetic cancellation with a rate of $10^{15}$\,Mx\,s$^{-1}$ is found. The released magnetic energy is redistributed by the siphon flows. The second group corresponds to two footpoints rooted in mixed-magnetic-polarity regions, where magnetic cancellation occurred at a rate of $10^{15}$\,Mx\,s$^{-1}$ and line profiles with enhanced wings of up to 200\,\kms\ were observed. These are suggestive of explosive-like events. The Doppler velocities combined with the SJ images suggest possible anti-parallel flows in finer loop strands. In the third group, interaction between two cool loop systems is observed. Evidence for magnetic reconnection between the two loop systems is reflected in the line profiles of explosive events, and a magnetic cancellation rate of $3\times10^{15}$\,Mx\,s$^{-1}$ observed in the corresponding area. The IRIS observations have thus opened a new window of opportunity for in-depth investigations of cool transition region loops. Further numerical experiments are crucial for understanding their physics and their role in the coronal heating processes.
\end{abstract}


\maketitle

\section{Introduction}
\label{sect_intro}
The magnetized solar upper atmosphere is structured by numerous types of loops\,\citep{2003A&A...406.1089D,2014SSRv..tmp...52F}. 
Based on their temperatures, they are categorized into cool ($10^5-10^6$\,K), 
    warm ($1-2\times10^6$\,K) and hot ($\geqslant 2\times10^6$\,K) loops \,\citep{2014LRSP...11....4R}. 
Loops cooler than $10^5$\,K were also suggested to have a major contribution to
    the EUV output of the solar transition region\,\citep{1983ApJ...275..367F,1986SoPh..105...35D,1987ApJ...320..426F,1993ApJ...411..406D,1998ApJ...507..974F,2001ApJ...558..423F,2012A&A...537A.150S}.
They were also observed in CDS and SUMER off-limb observations \,\citep[][]{1997SoPh..175..511B,2000ApJ...533..535C}. 
Hotter and cooler loops tend not to be co-spatial\,\citep[e.g.][]{1997SoPh..175..487F,2000A&A...359..716S}, 
  though there are exceptions\,\citep[e.g.][]{1998SoPh..182...73K}.

\par
Loop heating is a major part of the great coronal heating problem that still remains an unresolved puzzle\,\citep{2006SoPh..234...41K, 2014LRSP...11....4R}. 
Two mechanisms have been proposed\,\citep[for details, see][]{2006SoPh..234...41K}. 
The first mechanism is the so-called ``steady heating'' and means  that loops are heated continuously. 
The second  suggested mechanism is the ``impulsive heating'' where loops are heated by low-frequency events, 
  e.g. nanoflare storms or resonant wave absorption\,\citep[][and references therein]{2006SoPh..234...41K}.
Plasma velocity is a key parameter that can help determine which heating mechanism is in operation\,\citep{2006SoPh..234...41K,2013ApJ...767..107W,2014LRSP...11....4R}. 
Uniform, steady heating results in a stationary temperature distribution and a balance between heat flux and radiation losses,
   therefore no strong flows would exist in the loop. 
In the cases of non-uniform steady heating and impulsive heating, 
   stronger plasma flows along the loop can be found due to imbalance of plasma pressure
   between the two loop footpoints. 
Which one of these heating mechanisms is at work mainly depends on the plasma and magnetic properties of the loops. 
It is widely accepted that warm loops are heated by impulsive processes\,\citep{2002ApJ...567L..89W,2003ApJ...593.1174W,2004ApJ...605..911C,2005ApJ...626..543W,2006SoPh..234...41K,2009ASPC..415..221K,2009ApJ...694.1256T,2009ApJ...695..642U}. However,  for hot loops, both impulsive\,\citep{2010ApJ...723..713T,2011ApJ...738...24V,2014ApJ...783...12U,2014ApJ...795...48C} and steady heating\,\citep{2010ApJ...711..228W,2011ApJ...740....2W} have been suggested.

\par
There exist only a limited number of studies on cool loops mainly due to instrumental  limitations.
By comparing observations and simulations, \citet{2006A&A...452.1075D} suggested that cool loops
   are heated by nonlinear heating pulses, i.e. transient events. 
As discussed above, flow properties can be used to infer the heating mechanism in a loop. 
They are mainly based on Doppler shift measurements that can only be derived from spectral data. 
Doppler shifts, however, carry only line-of-sight (LOS) information and may cancel out in data with insufficient resolution, 
   especially in cases of anti-parallel flows in close-by loop strands\,\citep[e.g.][]{2013ApJ...775L..32A}. 
Cool loops are difficult to identify in on-disk observations because of the strong background emission
   and/or LOS contamination by other features. 
\citet{2003A&A...406..323D} studied a loop-like bright feature observed in SOHO/CDS, 
   and found 21\,km\,s$^{-1}$ blue-shifted flow in the {\sc O\,v} lines  (T=$2.5\times10^5$\,K) 
   after correcting for the projection effect. 
By combining TRACE images and SUMER spectral observations, \citet{2006A&A...452.1075D} found a $\sim 20$\,km\,s$^{-1}$ red-shifted velocity
   in a footpoint of a cool loop in the SUMER {\sc N\,v} line (T=$2\times10^5$\,K). 
\citet{1997SoPh..175..511B} reported a $\sim 50$\,\kms\ LOS velocity along a loop
    observed off-limb in CDS data. 
\citet{2000ApJ...533..535C} studied off-limb cool loops in an active region. 
They reported that the Doppler velocities vary in different loops from 20\,\kms\ (red-shift) to $-$15\,\kms\ (blue-shift) 
   measured in SUMER {\sc O\,vi}\,(T=$3.2\times10^5$\,K) and Hydrogen Ly$\beta$\,(T=$1\times10^4$\,K). 
\citet{2004A&A...427.1065T} found supersonic flows ($\gtrsim 100$\,km\,s$^{-1}$) in a loop-like feature
   measured via the secondary Gaussian components of the SUMER {\sc O\,vi} lines.

\par
Now the \textit{Interface Region Imaging Spectrograph}\,\citep[IRIS,][]{2014SoPh..tmp...25D} monitors the solar
   transition region {with} unprecedented spatial and spectral resolution, 
   offering a great opportunity for in-depth investigations of cool solar loops. 
In the present work, we analyze an IRIS dataset taken in a near-disc-center active region. 
In the observations, a few clusters of loop strands are identified in the transition region \siiv\ lines (T$=6.3\times10^4$\,K).
In the past, the identification of individual strands in such fine-scale transition region loops
   was close to impossible because of the lower spatial resolution of the previous spectrometers, e.g. SUMER and CDS on board SoHO.
We, therefore, have now a unique opportunity to study in great detail plasma velocities, dynamics, evolution and interaction
   of cool loops
that is crucial for understanding the physics of loops forming in this temperature regime.
The article is organized as follow: we describe the observations in Section\,\ref{sect_obs}, 
   the wavelength calibration in Section\,\ref{sect_wave_cal}, 
   the results and discussions in Section\,\ref{sect_res} 
   and the conclusions in Section\,\ref{sect_con}.


\section{Observations}
\label{sect_obs}

\begin{figure*}[!ht]
\includegraphics[width=17cm,clip,trim=0.5cm 0cm 6.6cm 0cm]{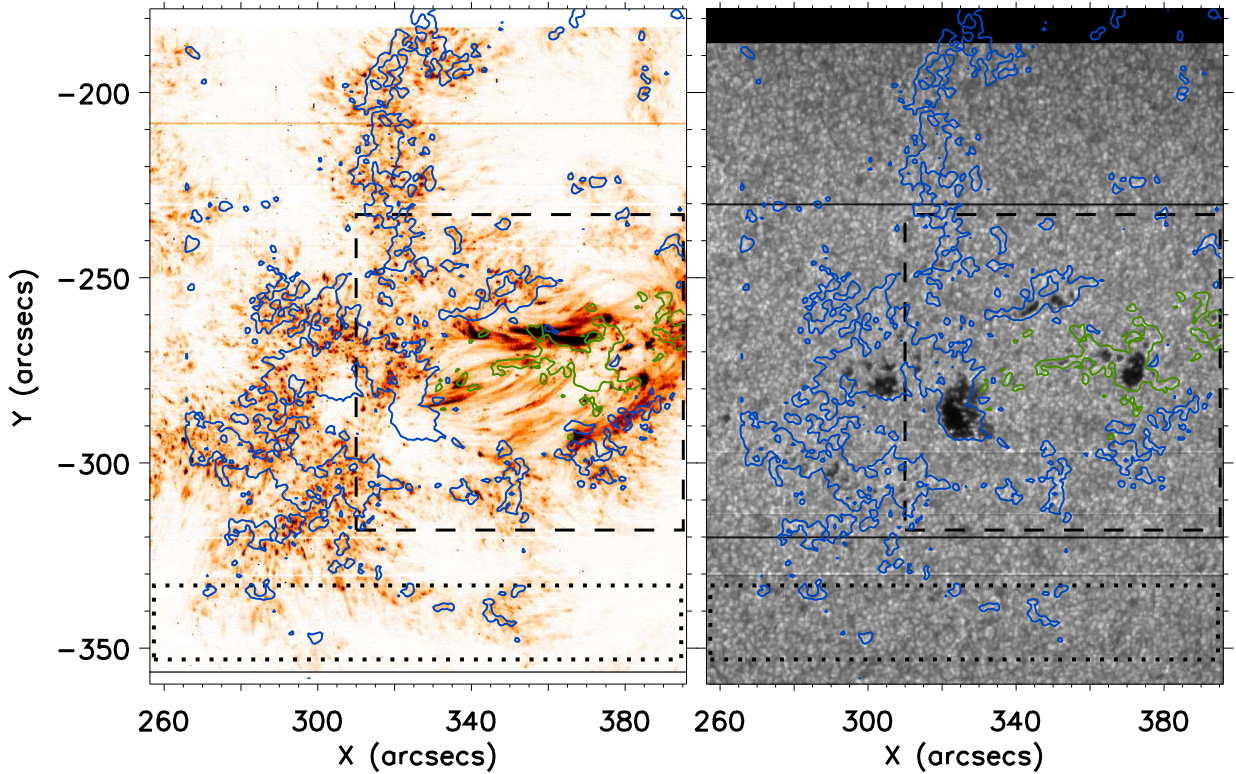}
\caption{IRIS raster in \siiv\,1402.8\,\AA\,(left, in reversed color table) and a 2832\,\AA\ continuum (right). The dashed lines indicate the region filled by clusters of cool transition region loops, while the dotted lines denote the region where the rest wavelength of \si\,1401.5\,\AA\ is determined. The blue and green contours represent the HMI longitudinal magnetic flux density levels at $-$200\,Mx\,cm$^{-2}$ and 200\,Mx\,cm$^{-2}$ respectively. \label{fig_obs_fov}}
\end{figure*}

\begin{figure*}[!ht]
\includegraphics[width=17cm,clip,trim=0.5cm 0.2cm 0cm 0cm]{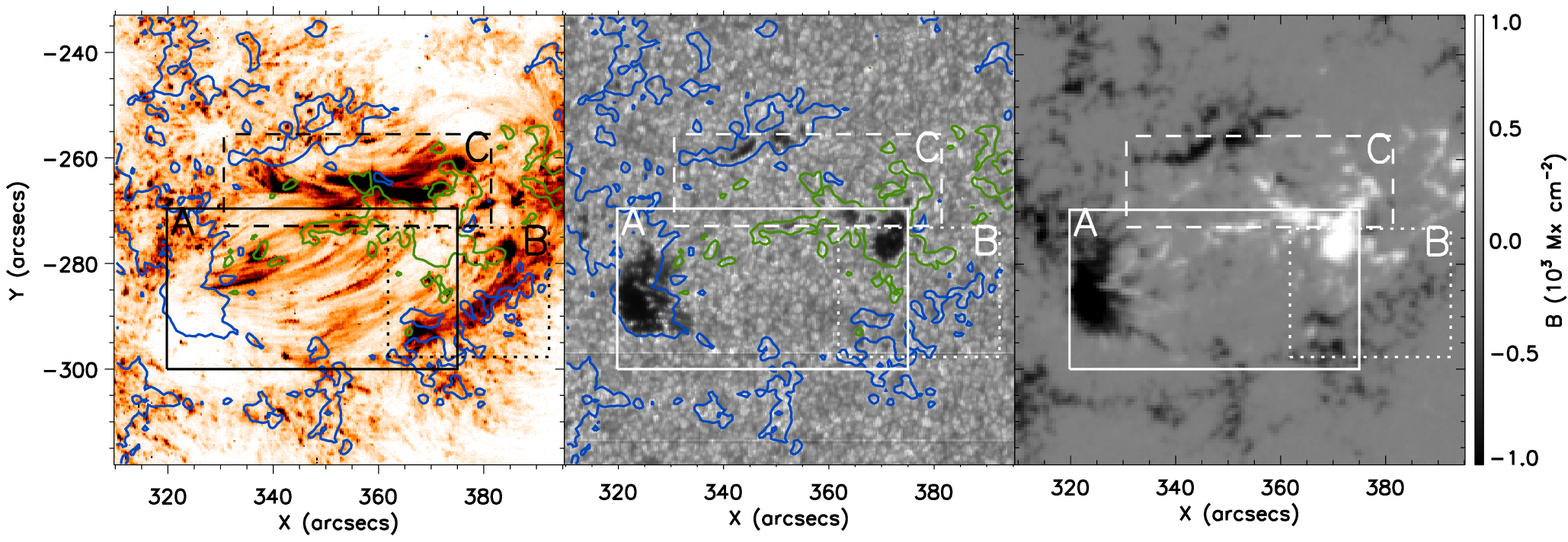}
\caption{Loop region viewed in IRIS \siiv\,1402.88\,\AA\,(left, in reversed color table), a 2832\,\AA\ continuum image (middle), and a HMI longitudinal magnetogram (right). The contours on the IRIS raster images correspond to HMI magnetic flux densities at $-$200\,Mx\,cm$^{-2}$ (blue) and 200\,Mx\,cm$^{-2}$ (green). ``A'' (with solid lines), ``B'' (with dotted lines) and ``C'' (with dashed lines) denote the three loop regions. \label{fig_lpfov_si}}
\end{figure*}

The IRIS observations were taken from 21:02\,UT to 21:36\,UT on 2013 December 27 targeting active region NOAA 11934. The spectral data were obtained by scanning the region from east to west in 400 steps with 0.35\arcsec step-size and 4\,s exposure time. The pixel size along the slit is 0.17\arcsec.  The field-of-view of the spectral data is 140\arcsec$\times$182\arcsec. The slit-jaw (SJ) images were recorded with a $\sim$10\,s cadence and a pixel size of $0.17\times0.17$\,arcsec$^2$. The IRIS data analysed here are level-2 products, on which dark current removal, flat-field, geometrical distortion, orbital and thermal drift corrections have been applied. A spatial offset between images taken in different spectral lines and SJ images found in \citet{2014ApJ...797...88H} is not present in this dataset. The new data production pipeline used after 2014 May should have performed well to handle all the calibration processes. In the present study, the raster scan using \siiv\,1402.8\,\AA\ and the SJ images taken in the 1400\,\AA\ channel are used to study cool transition region loops found in the observed FOV. These loops are also visible in the \cii\ raster. However, \cii\ shows self-absorption well seen now at the IRIS high spectral resolution\,\citep[see also e.g.][]{2014ApJ...797...88H}, therefore it is not suitable for calculations of plasma parameters.

\par
Figure\,\ref{fig_obs_fov} displays the radiance images of the IRIS spectral raster in the Si\,{\sc iv}\,1402.8\,\AA\ line and  the continuum emission summed from 2831.74\,\AA\ to 2833.27\,\AA. This figure gives an overview of the solar features (cool loops, sunspots and pores etc.) observed in this region. A sub-region filled by cool transition region loops located on the right hand side of the field-of-view is seen in the \siiv\ image. A few sunspots and pores connecting the loop systems are visible in the continuum image. 

\par
Line-of-sight magnetograms taken by the Helioseismic and Magnetic Imager\,\citep[HMI,][]{2012SoPh..275..229S} with a 45\,s cadence are also used in this study to determine the topology and evolution of the magnetic field of the region.

\section{Wavelength calibration}
\label{sect_wave_cal}
In order to measure velocities in the transition region \siiv\,1402.8\,\AA\ line, 
   the rest wavelength was determined first.
We applied an approach similar to the one proposed by \citet{2012ApJ...744...14Y} for Hinode/EIS hot lines. 
We first obtained the wavelength offset of \siiv\ 1402.8\,\AA\ relative to \si\ 1401.5\AA\ averaged from
   a quiet-Sun (QS) dataset taken from 2013 October 9 at 23:26\,UT to 2013 October 10 at 02:56\,UT. 
We note that the QS dataset has been used in a study by \citet{2014Sci...346A.315T} showing excellent quality. 
Doppler-shifts of neutral lines are normally considered to be close to zero in the QS\,\citep[see e.g.][]{1997SoPh..175..349B}. 
The field-of-view of the QS dataset covers various QS features, thus the average profile represents well the QS general characteristics. 
Next, an average profile of a quiet region in our dataset (marked by dotted lines in Figure\,\ref{fig_obs_fov}) was produced
   to determine the \si\,1401.5\,\AA\ line center. 
The quiet region is located away from the sunspots and covers a relatively large area (120\arcsec$\times$10\arcsec). 
Thus the \si\,1401.5\,\AA\ can be considered as being at rest, i.e. at zero Doppler shift. 
The obtained \si\,1401.5\,\AA\ line center with the offset given in the previous step added 
   is then considered as the rest wavelength of \siiv\,1402.8\,\AA. 
In this dataset, the rest wavelength of the line is found to be 1402.79\,\AA.

\par
When calculating nonthermal velocities, one needs to know the ion formation temperature
   and the line broadening caused by instrumental effects\,\citep[for details of the derivation, see][]{1998ApJ...505..957C}. 
In this study, we adopted the formation temperature given in the CHIANTI database\,\citep[V7.1.3,][]{1997A&AS..125..149D,2013ApJ...763...86L}, i.e. 6.3$\times$10$^4$\,K. 
For instrumental broadening, we used a method that is commonly applied to SUMER data\,\citep[see detailed description in][and references therein]{1998ApJ...505..957C}. 
This method assumes that a spectral profile emitted by a neutral atom is not broadened by any nonthermal motions. 
Therefore,  the broadening of an observed neutral line subtracted by its thermal part 
   gives the instrumental broadening. 
In this study, we used the \si\,1401.5\,\AA\ line obtained from the QS dataset mentioned above for this purpose. 
The instrumental broadening is found to be 57.3\,m\AA\ at a full width at half maximum (FWHM). 
This value is relatively large compare to 31.8\,m\AA\ given by the pre-flight measurements (see IRIS data user guide available on the web\footnote{http://iris.lmsal.com}). 
However, it is a practical approach to obtain the instrumental broadening during the flight.

\section{Results and discussion}
\label{sect_res}

In Figure\,\ref{fig_lpfov_si}, we show the IRIS \siiv\,1402.8\,\AA, the continuum and the HMI magnetogram images of the loop region. 
Clusters of cool transition region loops can be clearly seen in the \siiv\,1402.8\,\AA\ raster image (top-left panel). 
With the further aid of the 1400\,\AA\ SJ image sequence, 
   we identified three groups of loops that evolve differently. 
They are denoted as A, B and C (Figure\,\ref{fig_lpfov_si}). 
We describe their physical properties, dynamics and evolution in great detail in the following sections.

\subsection{Group A: cool transition region loop with one active footpoint}

\begin{figure*}[!ht]
\includegraphics[width=17cm,clip,trim=0.5cm 0.2cm 0cm 0cm]{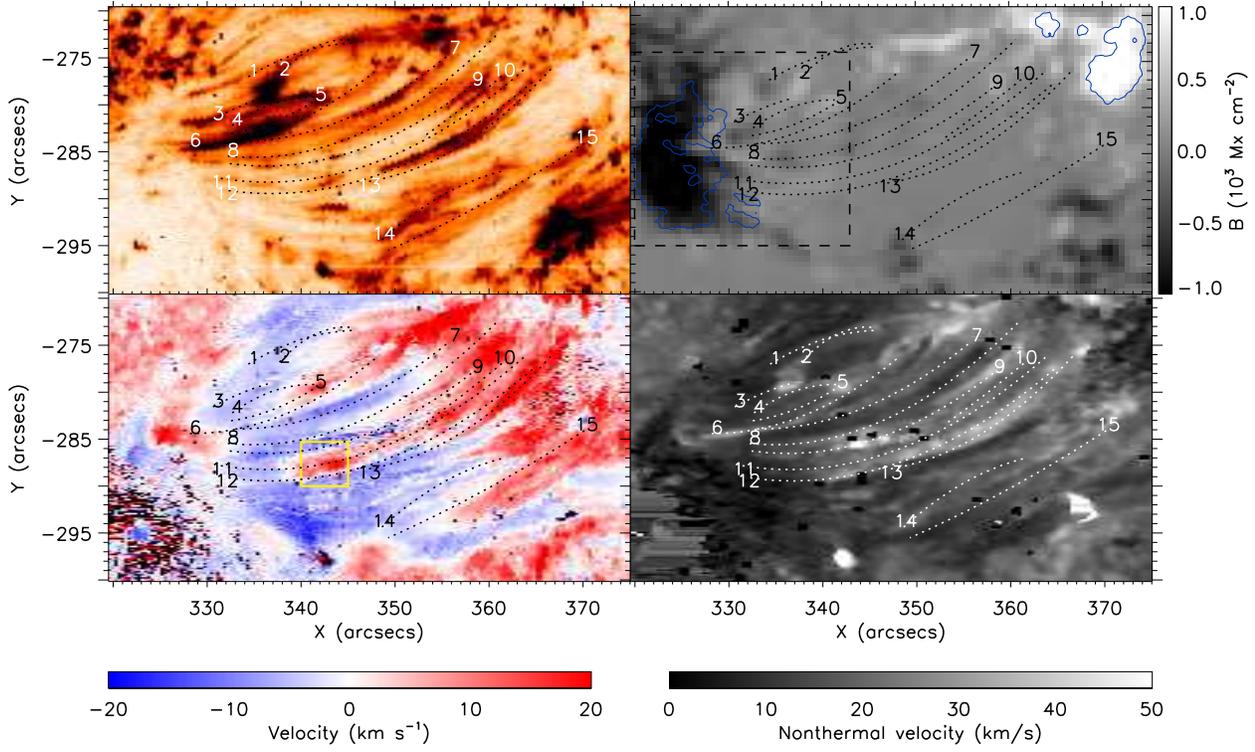}
\caption{Radiance image in \siiv\,1402.8\,\AA\ (top-left, in reversed color table), Doppler velocity image (bottom-left), nonthermal velocity image (bottom-right) and HMI magnetogram (top-right) of region ``A''. The dotted lines mark 15 loops that were visually identified. The blue contours over-plotted on the magnetogram denote the emission of the IRIS 2832\,\AA\ continuum raster image. The yellow box on the Doppler image outlines the region where anti-parallel flows are seen. The box (dashed lines) on the magnetogram marks the area from which the magnetic cancellation rate is obtained.\label{fig_lpa}}
\end{figure*}

\begin{figure*}[!ht]
\includegraphics[width=8.5cm,clip,trim=0cm 0.2cm 0.2cm 0cm]{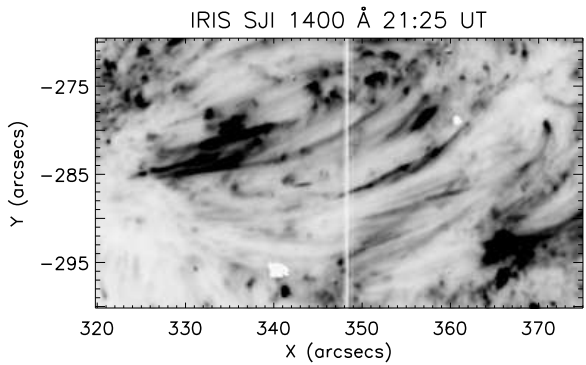}
\caption{IRIS SJI 1400\,\AA\ snapshot of region ``A''. The image is displayed in reversed color table. The white vertical line denotes the location of the spectrograph slit. (An animation is given online). \label{fig_rega_sj}}
\end{figure*}

\begin{figure*}
\includegraphics[width=8.5cm,clip,trim=0cm 0.2cm 0.2cm 0cm]{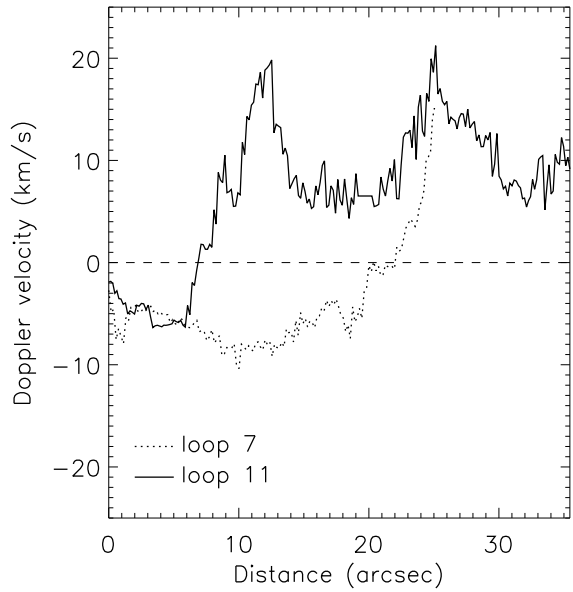}
\caption{Doppler velocity variations along loops 7 (dotted line) and 11 (solid line) starting from their southern legs. The dashed line denotes a zero Doppler shift. \label{fig_lpv_cur}}
\end{figure*}

\begin{figure*}
\includegraphics[width=17cm,clip,trim=0cm 0.2cm 0.1cm 0cm]{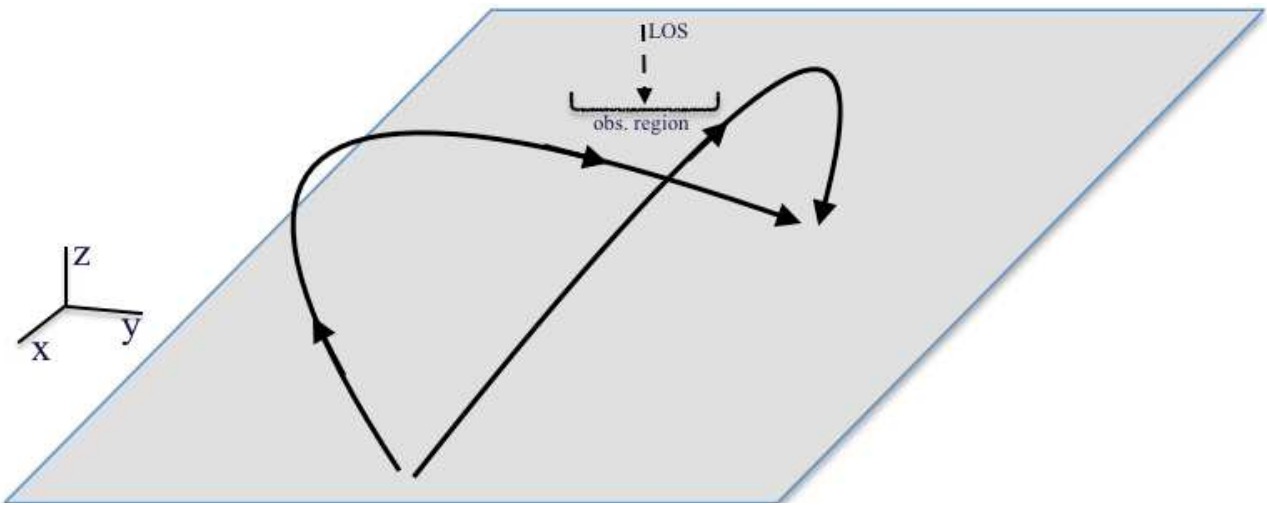}
\caption{Configuration of two loops that can produce locally oppositely directed Doppler shifts when observed along the `LOS' direction as shown in the figure. \label{fig_lp_geo}}
\end{figure*}

\begin{figure*}
\includegraphics[width=8.5cm,clip,trim=0cm 0.2cm 0.2cm 0cm]{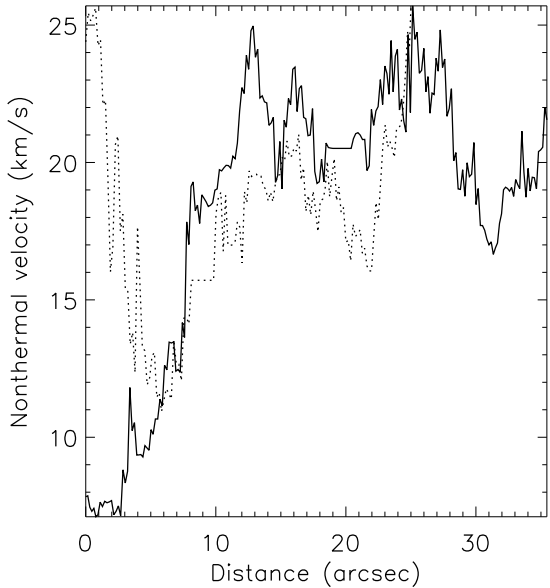}
\caption{Variation of nonthermal velocities along loops 7 (dotted line) and 11 (solid line) starting from their southern legs.\label{fig_lpw_cur}}
\end{figure*}

\begin{figure*}[!ht]
\includegraphics[width=9cm,clip,trim=0cm 0.2cm 0.2cm 0cm]{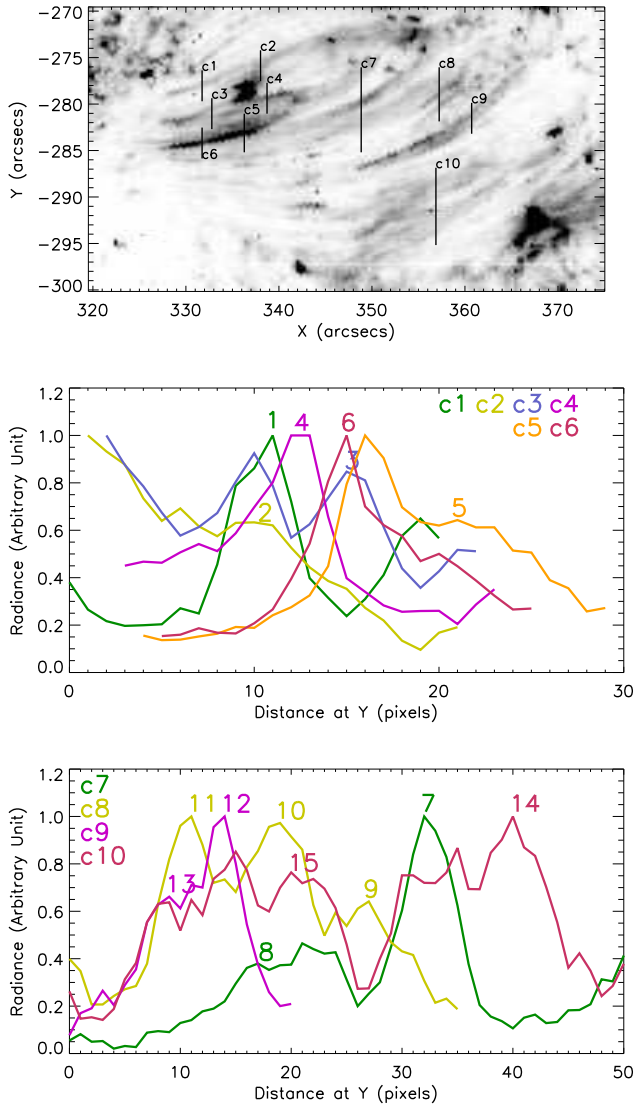}
\caption{Lightcurves (middle and bottom panels) along the cuts given on the IRIS\,\siiv\,1402.8\,\AA\ radiance image (black vertical lines, c1-- c10 in the top panel) crossing 15 visually-identified loops. Peaks denoted with numbers mark loop locations.\label{fig_lp_sjcross}}
\end{figure*}

The FOV containing group A is enlarged in Figure\,\ref{fig_lpa}. 
In this region, we visually identified 15 loops (see Figure\,\ref{fig_lpa}) using the \siiv\ radiance image and the SJ images (see Figure\,\ref{fig_rega_sj} and the online animation). The identification is based on cross-checking both the \siiv\ radiance image and the SJ images. Please note that this region is occupied by bundles of loops, many of which are clearly visible in the SJ images, though relatively weak in the \siiv\ radiance image (e.g. loops in the area below loop 12).
Most of these loops do not show clear footpoints in the spectral data. 
In the IRIS SJ images,  however, more loops with apparent plasma flows are visible (see online animation attached to Figure\,\ref{fig_rega_sj}). 
Most of these loops end in nearby a region of two small sunspots (Figure\,\ref{fig_lpa} top-right panel). 
They appear to be rooted in weak magnetic features spread around the strong sunspot fields. 
The southern ends of these loops are located in a mixed-magnetic polarity region, 
   while the northern ends are associated with a single polarity (positive) area. 
The apparent lengths of these loops vary from 10\arcsec ($\sim$7\,000\,km) to 40\arcsec ($\sim$30\,000\,km).

\par
The \siiv\ Dopplergram of the region (bottom-left panel of Figure\,\ref{fig_lpa}) clearly shows that
   the Doppler shifts gradually change from blue-shifted (negative values in the figure) in their southern legs
   to red-shifted (positive values in the figure) in their northern legs (except loop 6). 
This clearly indicates a plasma flow from the southern to northern legs of the loops. 
We note that a blue-shifted patch (red-shifted at the northern end) is present in the area between loop 12 and loop 15 where we did not identify any loops. This area is also occupied by loops that are clearly visible in the SJ images, and the blue/red-shifted patches should also represent siphon flows in loops in this region.
The blue-shifts are about 10\kms\ at the southern ends, while the red-shifts are about 20\kms\ at the northern ends.
Although the spectrometer slit scans the area only once, the Doppler velocities seem to be steady in the different loops
   scanned at different time. 
This indicates siphon flows in these loops that last at least for the duration of the observing period (i.e. $\sim$10 mins).

\par
We also noticed that the velocity distributions along individual loops are different. 
Blue-shifts dominate along the apparent length of some loops (e.g. loops 8, 11), 
   while red-shifts dominate in others (e.g. loops 7, 9 and 12). 
Figure\,\ref{fig_lpv_cur} gives the variation of the Doppler velocities along loops 7 and 11. 
We can see that about 90\% of the apparent length of loop 7 is blue-shifted, 
   while about 80\% of the apparent length of loop 11 is red-shifted. 
This velocity distribution leads to a localized region where ``anti-parallel flows'' are found
   (see e.g. the boxed region marked in the Dopplergram of Figure\,\ref{fig_lpa}). 
Local anti-parallel plasma flows have been reported in Hi-C observations by \citet{2013ApJ...775L..32A}, 
   who suggest plasma flows in a highway manner. 
We further compared the phenomenon found here with that in \citet{2013ApJ...775L..32A}. 
Our observation is based on Doppler velocity measurements that are different from the apparent velocities obtained in the Hi-C data.  In the present case, the oppositely directed Doppler velocities do not suggest plasma flows in opposite directions 
   as they converge to the same sign in the footpoints. 
In a localized region, they can simply result from the 3D geometry of the loops together with the projection effect. 
A possible geometry is drawn in Figure\,\ref{fig_lp_geo}.

\par
The nonthermal velocities of the region are shown in the bottom-right panel of Figure\,\ref{fig_lpa}. 
In general, the nonthermal velocities do not significantly vary in major part of the loops. 
Figure\,\ref{fig_lpw_cur} shows the variations of the nonthermal velocities along loops 7 and 11. 
Along loop 7, a dip is found at a distance of about 5\arcsec in the curve. 
After checking the selected loop path, we found that the first 5\arcsec of the loop seems to be affected
   by the activity occurring in loop 6. 
Appart from that, the nonthermal velocities of both loop 7 and loop 11 show a sharp  increase
   at the first few arcseconds near their southern footpoints, and then fluctuate from 15\,\kms\ to 25\,\kms. 
This implies that a dynamic process takes place in their southern legs where the energy and velocity of the plasma are gained. 
This is also supported by the mixed-polarity magnetic features in this footpoint region. 
Mixed-polarity magnetic features are usually associated with energetic events\,\citep[e.g. coronal bright points, explosive events, X-ray jets and flares,][and references therein]{2012A&A...548A..62H,huangzh2013,2014ApJ...797...88H}. 
Nonthermal velocities of $\sim$20\,\kms\ found in these loops are consistent with the typical values obtained in the transition-region
 SUMER data\,\citep{1998ApJ...505..957C}. 
We, therefore, suggest that the loops in this group A are likely to have only one active footpoint at its southern end that is responsible for the siphon flows.

\begin{figure*}[!ht]
\includegraphics[width=17cm,clip,trim=0.5cm 0.2cm 0.2cm 0cm]{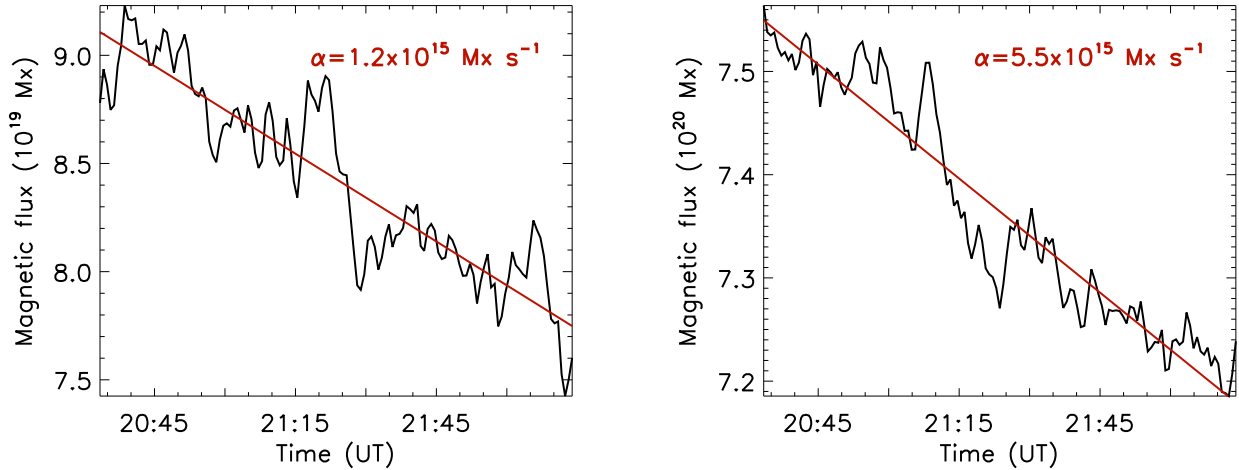}
\caption{Variation of the positive (left panel) and negative (right panel) magnetic fluxes obtained in the southern ends of the loops (marked by black dashed lines on the magnetogram shown in Figure\,\ref{fig_lpa}) in region ``A''. The red lines are the linear fits of the curves. The magnetic cancellation rates derived from the fittings are given by the $\alpha$ values. \label{fig_rega_cancel}}
\end{figure*}
    
\begin{figure*}
\includegraphics[width=17cm,clip,trim=0cm 1.5cm 0cm 0cm]{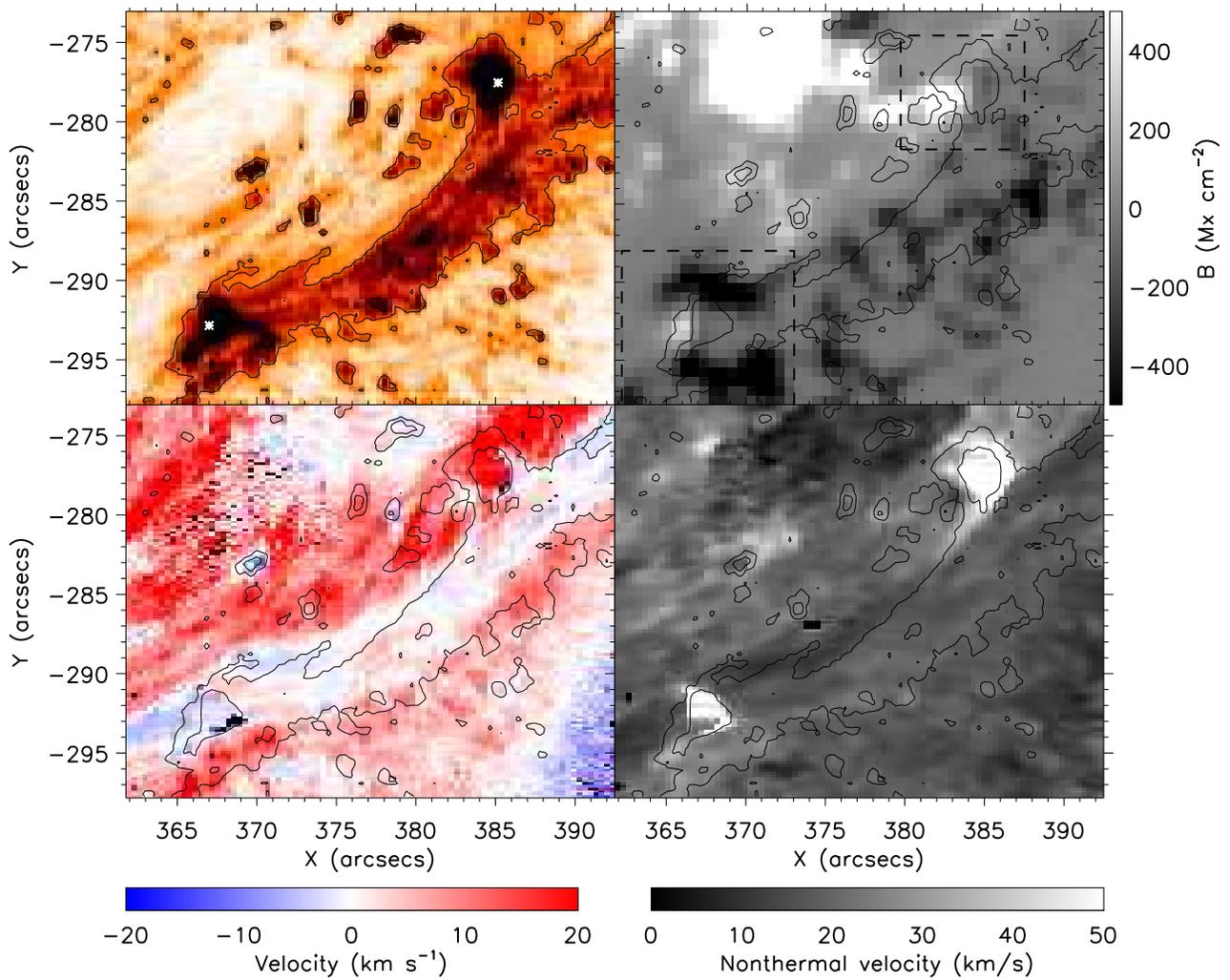}
\caption{Radiance image in \siiv\,1402.8\,\AA\  (top-left, in reversed color table), a Doppler velocity image (bottom-left), a nonthermal velocity image (bottom-right) and a HMI magnetogram (top-right) of region ``B''. The contours of the radiance image are over-plotted on all images. The loops inside the contours are better resolved in the SJ images shown in Figure\,\ref{fig_regb_sj}. The asterisks on the radiance image mark the places where the spectral lines presented in Figure\,\ref{fig_regb_line_smp} are taken from. The dashed lines on the magnetogram mark the area from which the magnetic cancellation rate is obtained.\label{fig_lpb}}
\end{figure*}

\begin{figure*}
\includegraphics[width=17cm,clip,trim=0cm 0cm 0cm 0cm]{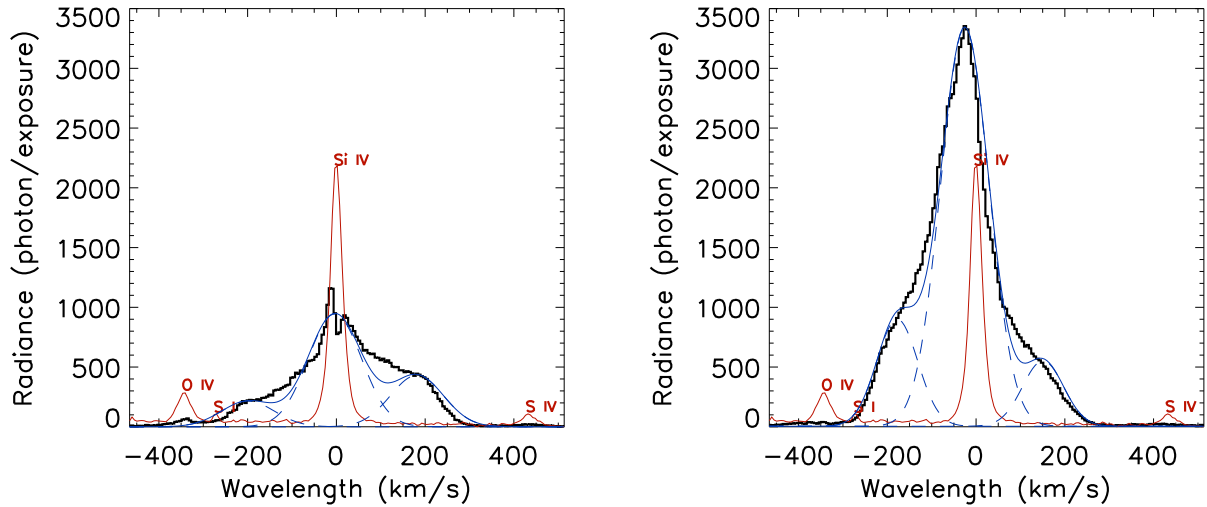}
\caption{Examples of the \siiv\,1402.8\,\AA\ profiles taken from the southern (left) and northern (right) footpoints of the loops in region B (asterisks in Figure\,\ref{fig_lpb}). The blue solid lines are the resulting triple Gaussian fit. The blue dashed lines are the three Gaussian components. The average profiles from the region described in Section\,\ref{sect_wave_cal} are shown in red (increased 200 times). The identified emission lines are marked. \label{fig_regb_line_smp}}
\end{figure*}

\begin{figure*}
\includegraphics[width=17cm,clip,trim=1cm 1.5cm 0cm 2cm]{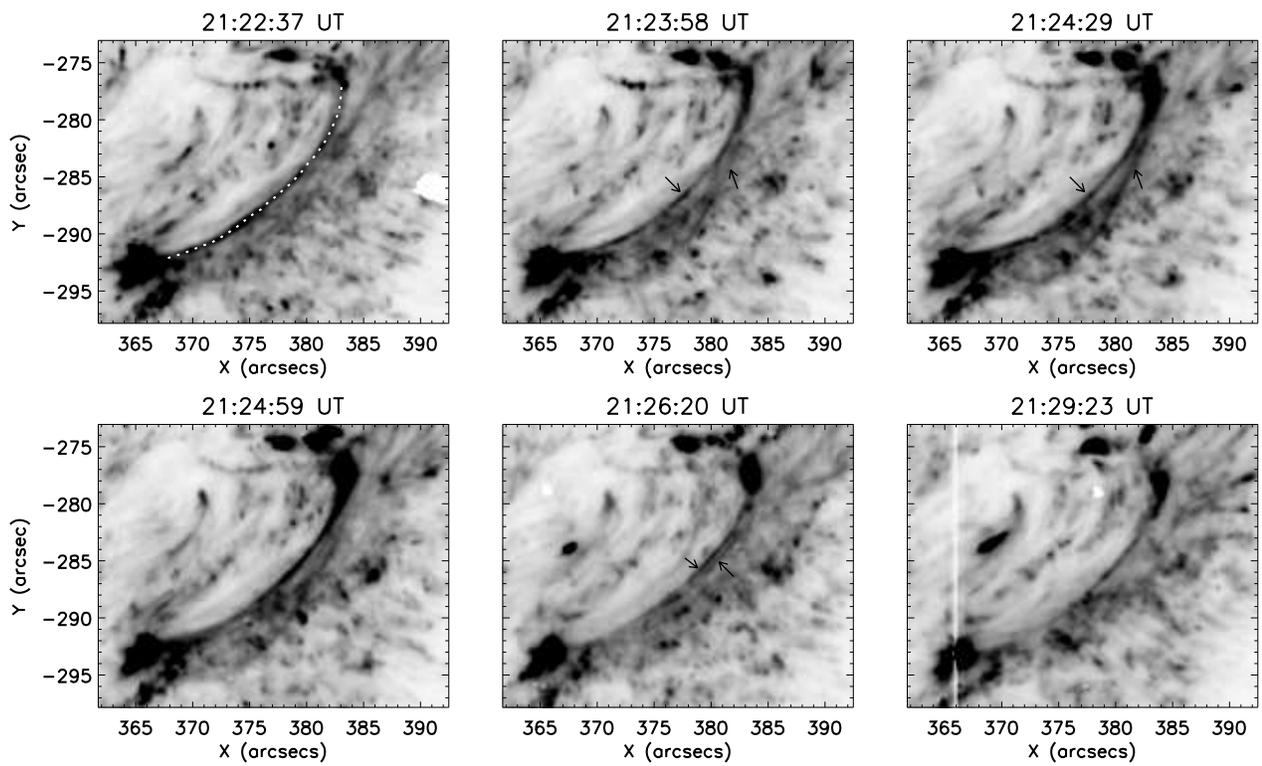}
\caption{Evolution of the loops in region B seen in IRIS SJ 1400\,\AA\, images (in reversed colour table). The dotted line in the image at 21:22:37\,UT outlines a loop. Arrows denote the locations of the loops in the region. The white vertical line in the image at 21:29:23\,UT is the IRIS spectrometer slit. (An animation is given online). \label{fig_regb_sj}}
\end{figure*}

\begin{figure*}
\includegraphics[width=17cm,clip,trim=0cm 0.5cm 0cm 0cm]{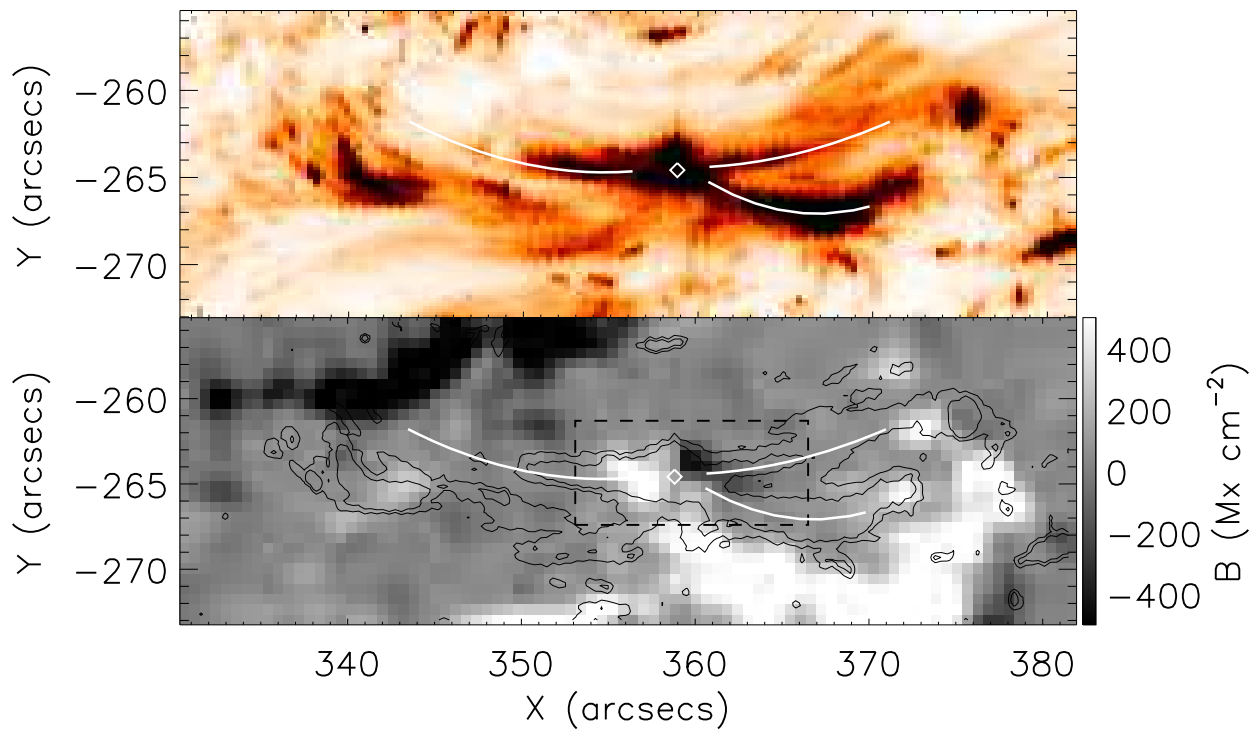}
\caption{Radiance image in \siiv\,1402.8\,\AA\ (top, in reversed color table) and a HMI magnetogram (bottom) of region ``C''. The contours of the radiance image are over-plotted on the magnetogram. The diamond symbol denotes the conjunction of the two loop systems. Three loops identified from the radiance image are marked by white solid lines. The dashed lines on the magnetogram mark the area from which the magnetic cancellation rate is obtained.\label{fig_lpc}}
\end{figure*}

\begin{figure*}
\includegraphics[width=17cm,clip,trim=1cm 1cm 0cm 0cm]{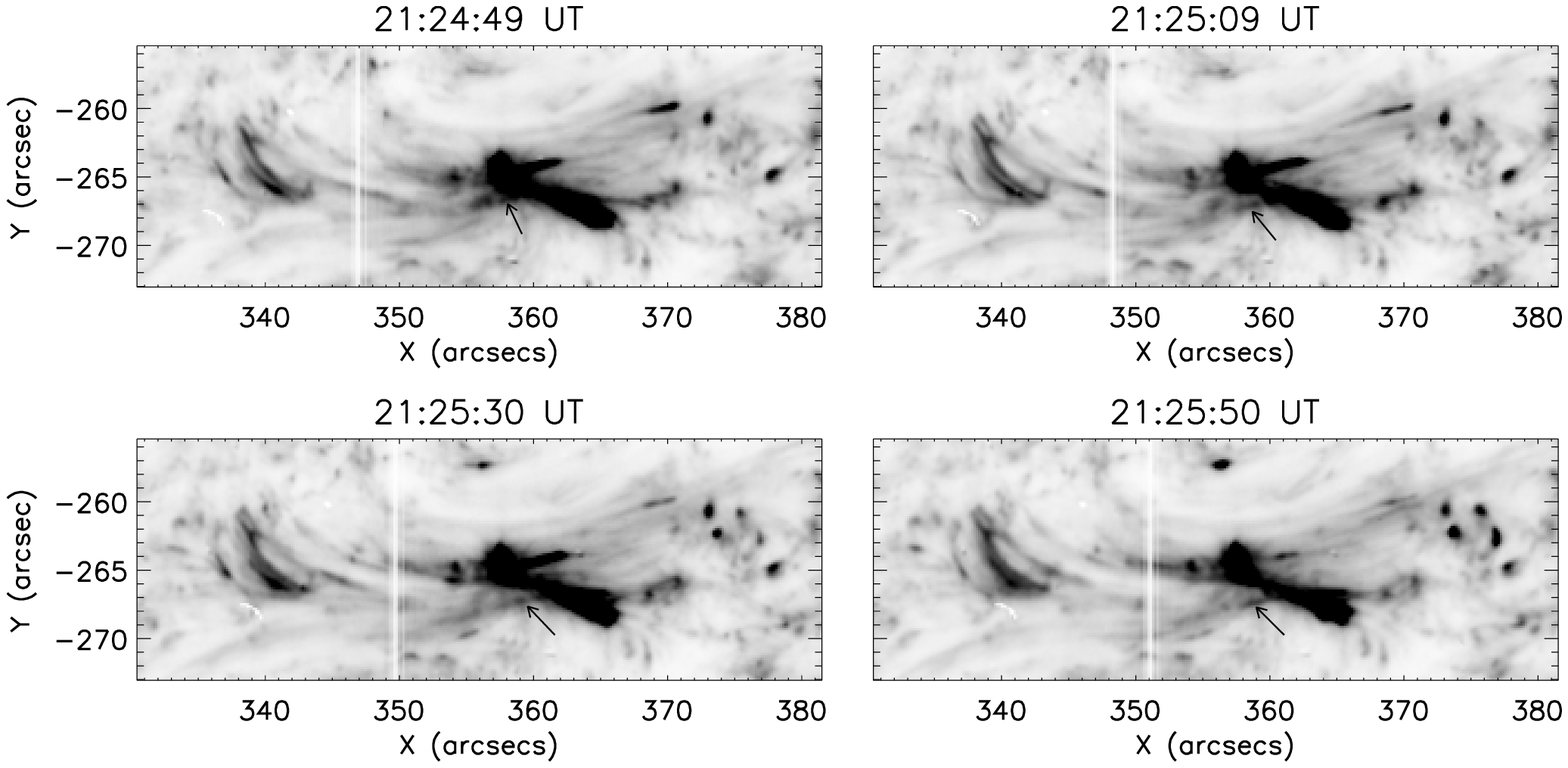}
\caption{Evolution of the loops in region C seen in the IRIS SJ 1400\,\AA\ images (in reversed colour table). The arrows denote the evolution of a loop moving away from the active conjunction. (An animation is given online). \label{fig_regc_sj}}
\end{figure*}

\par
The unprecedented high spatial resolution of the IRIS images gives us an opportunity to measure the cross-sections of these loops. 
In the past, the instrumental resolution was one of the main issues for such measurements\,\citep{2014LRSP...11....4R}. 
In Figure\,\ref{fig_lp_sjcross} we show the \siiv\,1402.8\,\AA\ emission along the cuts across the loops given in the top panel. 
The location of a loop is then determined by a peak in the lightcurve, while the dips at both sides determine the edges of the loop projection
   on the disk. 
If a loop is well isolated, its cross-section is then defined by the FWHM of the curve. 
One can see from Figure\,\ref{fig_lp_sjcross} that it is not possible to separate most of the loops from other features in the FOV. 
Therefore their cross-sections can not be measured due to the overlapping effect. 
From 15 loops, we could carry measurements for loops 1, 4 and 7 only. 
Their cross-sections are 4 pixels ($\sim$490\,km), 3.1 pixels ($\sim$382\,km), and 5.1 pixels ($\sim$626\,km), respectively. 
Given that the cuts are not perpendicular to the loop, the real cross-section should be less than the obtained values. 
Although the cross-sections of the rest of the loops can not be measured accurately, 
   a large number of them should have similar or even smaller sizes according to their appearances compared to loops 1, 4 and 7. 
In comparison with warm and hot loops measured by the \mbox{Hi-C} that are found to have cross-sections ranging from 300\,km\,\citep{2013ApJ...772L..19B} to 1\,000\,km\,\citep{2013A&A...556A.104P}, the cross sections measured in our data are close to lower limit of the Hi-C measurements. 
Loops forming at transition region temperatures seem to have finer scales than coronal loops.

\par
Spectroscopic measurements bring important information that can help to understand the physical mechanism
   that heats and drives plasma flows along cool loops.
We assume that the typical temperature of these loops
    is $\sim 6\times 10^4$~K,
    the loop length is $\sim 10^4$~km and the electron density is $\sim 10^{10}$~cm$^{-3}$ (a value representative
    of the transition region).
The observed siphon flow field lasts for at least 10\,mins, which is consistent with the picture where
    these loops are heated in a steady manner by an energy source in the blue-shifted footpoints (southern legs).
Given that there are an infinite number of ways for prescribing the spatial heating profile,
    we, therefore, choose to focus on one mechanism where the heating derives from Alfv\'en waves~\citep[for details, see][]{2005A&A...435.1159O}.
The wave energy, originally in the low-frequency Magnetohydrodynamic regime,
    is turbulently cascaded towards high-frequencies until proton-cyclotron resonance comes into play,
    whereby the energy of the turbulently generated proton-cyclotron waves
    are absorbed by protons.
Electrons are then heated via Coulomb collisions with protons.
    Assuming that the Alfv\'en waves are generated at locations in the chromosphere where the temperature
    is $2\times 10^4$~K, what is appealing in this mechanism is that $T_{M}$, the maximum
    temperature a loop acquires, depends only on the looplength $L$ and the imposed wave amplitude $\xi$.
With $\xi$ being constrained by SOHO/SUMER measurements~\citep[e.g.,][]{1998ApJ...505..957C}, 
    the lowest value that $T_M$ attains depends only on $L$.
An exhaustive parameter study~\citep{2005A&A...435.1159O} suggests that
    $T_{M}$ for a loop length of $\sim 10^4$~km always exceeds $0.7$~MK, which
    is much higher than the formation temperature of \siiv.
Lowering $\xi$ from its nominal value of $14$~km/s to $10$~km/s does not lower $T_M$ much.
    Introducing further complications such as the possibly unresolved magnetic twists actually makes $T_M$
    even higher~\citep{2006RSPTA.364..533L}. 
    We note that even within the class of models where Alfv\'en waves are the primary energy source for coronal heating,
   there are ways for dissipating the wave energy other than the parallel-cascade scenario adopted in \citet{2005A&A...435.1159O}.
For instance, \citet{2011ApJ...736....3V} proposed that the turbulent cascade may proceed primarily in the lateral rather than in the mean-field direction
   as a result of the nonlinear interactions between counter-propagating Alfv\'en waves.
While a definitive answer is still not available regarding how the loop temperatures and densities
   depend on loop lengths in this scenario, this loop-length dependence may
   well be different from what was found in \citet{2005A&A...435.1159O}.
In view of this, we conclude that the Alfv\'en wave heating mechanism described by \citet{2005A&A...435.1159O} is unlikely to reproduce 
   the measured loops that emit strongly in \siiv.
However, we cannot rule out the possibility that other steady heating mechanisms, such as the one by \citet{2011ApJ...736....3V},
   may provide a satisfactory explanation for the physical properties of this set of cool transition region loops observed by \textit{IRIS}.

\par
Another mechanism that may account for the velocity field is the impulsive heating that is often invoked in
    interpreting warm loop observations~(see Section\,\ref{sect_intro}).
To discuss this aspect further, we note that three timescales are relevant, namely
    the electron thermal conduction timescale $\tau_{\mathrm{cond}}$,
    radiative cooling timescale $\tau_{\mathrm{rad}}$,
   and the timescale $\tau_{\mathrm{enth}}$ at which enthalpy flux plays a role.
Let $n_8$, $T_6$, and $L_9$ denote the electron density in $10^8$~cm$^{-3}$,
    temperature in $10^6$~K, and loop length in $10^9$~cm, respectively.
We then take $n_8 = 100$, $T_6=0.06$ and $L_9 = 1$ to be representative of the observed cool loops.
With $\tau_{\mathrm{cond}} = 160 n_8 L_9^2/T_6^{5/2}$~\citep{2014LRSP...11....4R},
    one finds that $\tau_{\mathrm{cond}} \sim 10^7$~seconds.
Likewise, one finds that $\tau_{\mathrm{rad}}\approx 40/(n_8 T_6)$ at the temperature range of interest,
    yielding $\tau_{\mathrm{rad}}\approx 7$~seconds.
On the other hand, the enthalpy timescale may be approximated with the longitudinal sound transit time
    given that the flow speed is not too far from the sound speed ($140\sqrt{T_6}\approx 30$~km/s).
This yields that $\tau_{\mathrm{enth}} \approx 70 L_9/\sqrt{T_6} \approx 280$~seconds. 
As the most likely location where the heating pulses are generated are
    the southern ends,
it then follows from the evaluation of the timescales that 
    a single heating pulse cannot account for the loop measurements. Neither the electron heat conduction
    nor the enthalpy is able to redistribute the heat along the loop because
    their timescales are substantially longer than the radiative cooling time scale.
Therefore a series of heating pulses is the most likely mechanism, and
    the total duration of these heating events should last longer than the enthalpy timescale (i.e. $\sim 5$~mins),
    and the temporal separation between two pulses should be substantially shorter than the
    radiative cooling timescale (i.e. $\sim 7$~secs).
    
\par
One may now ask what constitutes a heating pulse. As discussed in~\citet{2006SoPh..234...41K} and \citet{2012RSPTA.370.3217P}, this pulse
can be either AC (resonant wave absorption) or DC (nanoflares) in origin.

\par
Resonant wave absorption is reasonable in this area because the loops are rooted in a sunspot area 
    where masses of different modes of oscillations have been suggested and reported\,\citep[see e.g.][and references therein]{Marsch2004,2014ApJ...792...41Y}. 
By applying a wavelet analysis to various locations in these loops, we find no signature of oscillations. 
However, the wave scenario can not be ruled out because observational effects (such as overlapping, low temporal resolution)
    can be the reason that no periodicity is found in our analyses.

\par
Nanoflares were first proposed by \citet{1988ApJ...330..474P} as the most basic unit of impulsive energy release in the solar atmosphere. 
They are defined by the released total energy rather than any particular event. 
By definition the nanoflare energy is $\sim 10^{23}$\,erg, i.e. $10^9$ times smaller than the flare energy ($10^{32}$\,erg). 
It has been suggested that a nanoflare is the signature of small-scale magnetic reconnection occurring in the solar atmosphere\,\citep[see e.g.][]{2002ApJ...565.1298W,2013Natur.493..501C}. 
For these loops, the mixed-polarity magnetic features in the southern ends suggest that magnetic reconnection might take place in the area. 
We further investigated the magnetic flux in the southern footpoints. 
In Figure\,\ref{fig_rega_cancel}, we show the variation of the magnetic flux in this area from 20:30\,UT to 22:13\,UT. 
Magnetic cancellation is clearly present with rates of $1.2\times10^{15}$\,Mx\,s$^{-1}$ for the positive flux and $5.5\times10^{15}$\,Mx\,s$^{-1}$ for the negative flux. The cancellation rate for the negative flux is larger because the calculation includes the large sunspot that might cancel with the positive flux outside the selected box. High magnetic cancellation rates have been widely related to various energetic events, e.g. flares, jets, explosive events, etc.
For comparison, the cancellation rates are about $8\times10^{15}$\,Mx\,s$^{-1}$ in footpoints of an active region outflow\,\citep{vanninathan2015}, $5\times10^{14}$\,Mx\,s$^{-1}$ in an explosive event\,\citep{2014ApJ...797...88H} and X-ray jets\,\citep{2012A&A...548A..62H}, and $10^{16}$\,Mx\,s$^{-1}$ in a GOES C4.3 flare\,\citep{huangzh2013}. 
Therefore, the $10^{15}$\,Mx\,s$^{-1}$ cancellation rate suggests a significant magnetic energy release in the footpoints of these loops. 
Since the cancellation rate here is obtained from a large area that should include multiple energetic events, 
   energy released in a single event might be less than that in one explosive event\,\citep[i.e. $\sim10^{24}$ erg, see e.g.][]{2002ApJ...565.1298W}. 
Unfortunately, the current data do not allow a quantitive analysis of the energy release in the region. An investigation using high-resolution vector magnetic field data and/or simulations could provide a deeper insight on the issue.

\par
To conclude, the cool loops in group A are possibly heated by a series of 
    heating pulses in their southern legs produced by magnetic reconnection.
The magnetic reconnection scenario is strongly supported by the observed continuous magnetic flux cancellation.
The magnetic energy is released in an intermittent manner with a cadence less than 7\,s.
The enthalpy flux associated with the sub-sonic flow plays an important role in 
    constantly redistributing the heat deposited therein.

\subsection{Group B: cool transition region loop with two active footpoints}
Group B includes loops with compact brightenings in both footpoints (top-left panel of Figure\,\ref{fig_lpb}, where the black patches marked by asterisk symbols are the footpoints of the loops), 
   which are located in mixed-polarity magnetic flux regions (see top-right panel of Figure\,\ref{fig_lpb}). 
The Doppler shifts (bottom-left panel of Figure\,\ref{fig_lpb}) along these loops are very different from those in group A. 
No signature of siphon flows is recorded in the Doppler velocities. 
Instead, blue and red Doppler shifts ranging from $-5$\,\kms\ to 12\,\kms\ are distributed alternately along the loops. 
The nonthermal velocities of the region (bottom-right panel of Figure\,\ref{fig_lpb}) clearly show large values in the footpoints 
   where explosive-event line profiles are observed (see the discussion below). 
The nonthermal velocities are in the range from 10\,\kms\ to 25\,\kms\ with higher values at the two ends of the loops.

\par
The footpoint regions of the loops are very dynamic, strongly suggesting that the energy sources for the loop heating are located there. 
We further investigated the \siiv\,1402.8\,\AA\ line profiles in the region, and we found typical explosive-event line profiles
   in the footpoints (see the examples given in Figure\,\ref{fig_regb_line_smp}). 
The line profiles suggest that bi-directional plasma flows with velocities of about 200\,\kms\ are produced. 
This type of line profiles has been suggested to be a signature of magnetic reconnection in the transition region\,\citep[e.g.][among others]{1997Natur.386..811,2014ApJ...797...88H,2014Sci...346C.315P}. 
Magnetic cancellation is also found in the two footpoints of the loop group. 
The cancellation rates derived from the two boxed regions outlined in Figure\,\ref{fig_lpb} are found to be about $10^{15}$\,Mx\,s$^{-1}$ 
    indicating the release of a large amount of magnetic energy.

\par
Explosive-event line profiles in both footpoints also imply that oppositely directed flows might take place. 
The small Doppler shifts found in the loops away from the footpoints might 
    result from oppositely directed flows being canceled out in nearby loop strands.
To investigate this possibility,
   we trace the temporal evolution of the loops seen in the IRIS SJ 1400\,\AA\ images (see online animation, Figure\,\ref{fig_regb_sj}). 
We found that the group consists of multiple fine loop strands  (see Figure\,\ref{fig_regb_sj}, marked by arrows). 
Their footpoints are rooted in compact small areas. 
Two parallel loops are clearly seen at 21:26:20\,UT (indicated with arrows). 
This again supports the possibility that anti-parallel flows in close-by loop strands cancel out to produce the detected Doppler velocities in the raster scan. 
We would like to note that the SJ images have higher spatial resolution in the solar X direction (0.17\, arcsec/pixel) 
   than the raster scan that is defined by the slit width (0.35 arcsec/pixel). 
Blue and red Doppler shifts appear along the loops suggesting that flows with one and the other direction are present.

\subsection{Group C: interactions of two cool transition region loop systems}
Group C consists of two loop systems, which are interacting with each other, and therefore show very dynamic evolution. 
The interaction occurs at the conjunction of the two of the footpoints of the loop systems (see the diamond symbol in Figure\,\ref{fig_lpc}). 
The \siiv\,1402.8\,\AA\ profiles in most pixels of the region are non-Gaussian. 
Explosive-event line profiles are seen all over in this area. 
This is also supported by the observed high rate of magnetic cancellation of about $3\times10^{15}$\,Mx\,s$^{-1}$. 
Magnetic reconnection possibly occurs between the two interacting loop systems. 
This kind of reconnection should generate a small-scale loop in the reconnection site which then submerges and a long loop connecting the two far ends. 
To further investigate this,  we analyzed the temporal evolution of the region in the IRIS SJ 1400\,\AA\, images (see Figure\,\ref{fig_regc_sj} 
   and the attached animation). 
We observed a loop rising from the conjunction region (see the arrows in Figure\,\ref{fig_regc_sj}), 
   which is most likely one of the long loops formed during the reconnection process. 
This loop, however, disappears after 21:26\,UT. 
It might have been heated to higher temperatures or the plasma has drained towards the footpoints. 
We also searched for signatures of heated loops in the 171\,\AA\ channel of the Atmospheric Imaging Assembly\,\citep[AIA,][]{2012SoPh..275...17L}, 
   where the conjunction region is clearly visible. 
Because of the low resolution, we see some fuzzy cloud-like features rising from the conjunction area rather than any clear loop structures. 
A coronal imager with higher spatial resolution is essential to obtain a firm answer on this issue.

\section{Summary and conclusions}
\label{sect_con}
In the present work, we have studied clusters of cool transition region loops in NOAA AR11934 observed by IRIS from  21:02\,UT to 21:36\,UT on 2013 December 27 
   at unprecedented resolution in the \siiv\,1402.8\,\AA\ line. 
The analysed cool transition region loops are finely structured and have different dynamics and evolution. 
Three groups of loops were identified and studied in great detail.

\par
The loops in group A are relatively stable. They are finely structured with cross-sections of about 382--626\,km. 
Their southern legs are rooted in a mixed-magnetic polarity region near a small sunspot while their northern legs are associated with a single magnetic polarity. 
Possible siphon flows in these loops are suggested by the \siiv\,1402.8\,\AA\ Doppler velocities that are gradually changing from about 10\,\kms\ blue-shifts in the southern legs to about 20\,\kms\ red-shifts in the northern ones. 
The nonthermal velocities in the major sections of the loops vary from 15\,\kms\ to 25\,\kms, but increase in the southern ends. 
We concluded that these loops can not be heated by a steady energy release process and impulsive heating mechanism is required. 
The  energy is possibly deposited in their southern ends where magnetic cancellation with a rate of $10^{15}$\,Mx\,s$^{-1}$ indicates
   the release of significant magnetic energy. 
The magnetic energy is likely to be released impulsively by magnetic reconnection, and it is redistributed by the enthalpy flux carried by the siphon flows.

\par
The loops in group B have two active footpoints seen in \siiv\,1402.8\,\AA. 
Both footpoints are located in mixed-polarity regions. 
Small-scale magnetic reconnection events are found in the footpoints, which are witnessed by explosive-event line profiles
   with enhanced wings at about 200\,\kms\ Doppler shifts and magnetic cancellation with a rate of about $10^{15}$\,Mx\,s$^{-1}$. 
These loops are possibly impulsively powered by small magnetic reconnection events occurring in the transition region. 
Doppler velocities along the loops reveal that blue and red Doppler-shifts ranging from $-$5\,\kms\ to 12\,\kms\ alternate along the loop. 
The nonthermal velocities vary from 10\,\kms\ to 25\,\kms. 
These loops viewed in the SJ images show finer strands in which oppositely directed plasma flows might be present.

\par
Group C is an excellent example of the interactions of two loop systems that have two end points rooted in the same region. 
This conjunction region is associated with a mixed-polarity magnetic flux and is very dynamic as witnessed by strong non-Gaussian \siiv\ line profiles. 
Magnetic cancellation with a rate of $3\times10^{15}$\,Mx\,s$^{-1}$ is found in the area. 
Explosive-event line profiles suggest that magnetic reconnection occurs between these two loop systems.

\par
In summary, cool transition region loops differ a lot from warm and hot loops. 
They are finely structured, more dynamic and diverse. 
Impulsive heating involving magnetic reconnection is the most plausible heating mechanism. Future numerical studies are required to fully understand the physics of cool loops and their role in coronal heating.

\acknowledgments
{\it Acknowledgments:}
We would like to thank the anonymous referee for the helpful and constructive comments, and Dr. Hardi Peter for the useful discussion.
This research is supported by the China 973 program 2012CB825601, the National Natural Science Foundation of China under contracts: 41404135 (ZH), 41274178 and 41474150 (LX \& ZH), and 41174154, 41274176 and 41474149 (BL), the Shandong provincial Natural Science Foundation ZR2014DQ006 (ZH) and special funds from Shandong provincial postdoc innovation project 201402034 (ZH). MM is supported by the Leverhulme Trust. 
Research at the Armagh Observatory is grant-aided by the N. Ireland Department of Culture, Arts and Leisure. 
IRIS is a NASA small explorer mission developed and operated by LMSAL with mission operations executed at NASA Ames Research centre and major contributions to downlink communications funded by the Norwegian Space Center (NSC, Norway) through an ESA PRODEX contract. AIA and HMI data is courtesy of SDO (NASA). We thank JSOC for providing downlinks of the SDO data.

{\it Facilities:} \facility{IRIS}, \facility{SDO/HMI}.

\bibliographystyle{apj}
\bibliography{bibliography}

\end{document}